\documentclass[aps, pra, a4paper, twocolumn, noshowpacs, superscriptaddress, floatfix]{revtex4}
\usepackage{graphicx}
\usepackage{dcolumn}
\usepackage{color}
\usepackage{times}
\usepackage{bm}
\usepackage{amssymb}
\usepackage{amsmath}
\usepackage{epsfig}
\usepackage{epstopdf}
\usepackage{dsfont}
\usepackage{subfigure}
\begin{document}
\renewcommand{\thefootnote}{\fnsymbol {footnote}}
	
	\title{\textbf{Universal trade-off relation between coherence and intrinsic concurrence for two-qubit states}}
	
	\author{Xiao-Gang Fan}
	\affiliation{School of Physics \& Material Science, Anhui University, Hefei 230601, China}
	
	\author{Wen-Yang Sun}
	\affiliation{School of Physics \& Material Science, Anhui University, Hefei 230601, China}

    \author{Zhi-Yong Ding}
	\affiliation{School of Physics \& Material Science, Anhui University, Hefei 230601, China}

    \author{Fei Ming}
	\affiliation{School of Physics \& Material Science, Anhui University, Hefei 230601, China}

    \author{Huan Yang}
	\affiliation{School of Physics \& Material Science, Anhui University, Hefei 230601, China}
	
	\author{Dong Wang}
	\affiliation{School of Physics \& Material Science, Anhui University, Hefei 230601, China}
	\affiliation{CAS Key Laboratory of Quantum Information,University of Science and Technology of China, Hefei 230026, China}
	
	\author{Liu Ye}
	\email[Corresponding author: ]{yeliu@ahu.edu.cn}
	\affiliation{School of Physics \& Material Science, Anhui University, Hefei 230601, China}
	
\begin{abstract}
Abstract: Entanglement and coherence are two essential quantum resources for quantum information processing. A natural question arises of whether there are direct link between them. And by thinking about this question, we propose a new measure for quantum state that contains concurrence and is called intrinsic concurrence. Interestingly, we discover that the intrinsic concurrence is always complementary to coherence. Note that the intrinsic concurrence is related to the concurrence of a special pure state ensemble. In order to explain the trade-off relation more intuitively, we apply it in some composite systems composed by a single-qubit state coupling four typical noise channels with the aim at illustrating their mutual transformation relationship between their coherence and intrinsic concurrence. This unified trade-off relation will provide more flexibility in exploiting one resource to perform quantum tasks and also provide credible theoretical basis for the interconversion of the two important quantum resources.
\end{abstract}
\maketitle

\section{Introduction}
Entanglement and coherence are two crucial resources which are widely applied to quantum information processing and computation \cite{w01}. For a physical system, commonly used entanglement measures mainly consider correlations between their subsystems, whereas we usually think the physical system as a whole in the research of coherence omitting its structure \cite{w02}. Entanglement as one of earlier resource theories is crucial ingredient for various quantum information processing protocols \cite{w03}, such as remote state preparation \cite{w04,w05}, quantum teleportation \cite{w06}, super-dense coding \cite{w07} and so on. With the development of the entanglement theory, entanglement of formation \cite{w28}, concurrence \cite{w17}, relative entropy of entanglement \cite{w29}, negativity \cite{w30} have been proposed. Although entanglement can be measured in a variety of ways, there exist intrinsic relations between them. For instance, a functional relation between the entanglement of formation and concurrence has been put forward by Ref. \cite{w17}. As we all know that the negativity of a two-qubit state is invariably less than its concurrence, and its entanglement of formation is always greater its relative entropy of entanglement. On the other hand, coherence is usually a major concern of quantum optics in earlier research \cite{w08}. But a different viewpoint that quantum coherence was regarded as one of the key resources, just like entanglement, is proposed by Aberg \cite{w09} in 2006. Based on the work of Aberg, the resource theory of quantum coherence has been developed \cite{w10,w11,w12,w13,w14,w15}. The resource theory is based on the rules that the set of incoherent operations is seen as the free operations and the set of incoherent states is seen as the set of free states. These free operations and free states depend on a reference basis ${\left\{ {\left| n \right\rangle } \right\}_{n = 1,{\rm{ }} \cdots ,{\rm{ }}d}}$ of the d-dimensional Hilbert space. This means that the quantification of quantum coherence intrinsically rests with the reference basis.
Now that both entanglement and coherence are characterized by the resource theory, the understanding of common evolution of coherence and entanglement will be crucial. In particular, the research of the intrinsic relations hidden in these quantum resources has been made in recent years \cite{w31,w32}. Since the chosen types of quantum resources and measure approaches are various, there exist distinct differences for these intrinsic relations among the quantum states. Therefore, the main goal of our research is how to obtain a universal intrinsic relation.

In our research, we work out two puzzles. First, for a generally two-qubit mixed state, we will determine that its intrinsic concurrence can be complementary to its coherence. This reveals that the increase of its coherence causes the decrease of its intrinsic concurrence. Second, we will determine the internal relationship between its intrinsic concurrence and concurrence of a special pure state ensemble which is a special decomposition of a two-qubit state. This will solve the problem where do its coherence and concurrence come from. In Sec. II, we review the quantification of first-order coherence and concurrence. In Sec. III and Sec. IV, we provide detailed proofs of the complementary and the special decomposition respectively. In Sec. V, we illustrate its application for some typical systems.

\section{Preliminaries}
A commonly used entanglement measure is concurrence \cite{w03,w16}. For a two-qubit pure state $\left| \psi  \right\rangle $, its spin-flipped state is defined by $\left| {\tilde \psi } \right\rangle {\rm{ = (}}{\sigma _2} \otimes {\sigma _2})\left| {{\psi ^*}} \right\rangle $, where $\left| {{\psi ^*}} \right\rangle $ is the complex conjugate of $\left| \psi  \right\rangle $ , and ${\sigma _2}$ is the second one of the Pauli matrices. The concurrence is defined as \cite{w17}
\begin{align}
C\left( {\left| \psi  \right\rangle } \right) = \left| {\left\langle {\psi }
 \mathrel{\left | {\vphantom {\psi  {\tilde \psi }}}
 \right. \kern-\nulldelimiterspace}
 {{\tilde \psi }} \right\rangle } \right|.
\end{align}
For a general two-qubit state $\rho $, its density matrix can be written as the form $\rho  = \sum\limits_{n = 1}^4 {{p_n}\left| {{\psi _n}} \right\rangle \left\langle {{\psi _n}} \right|} $, where ${p_n}$ are the eigenvalues, in decreasing order, of the matrix $\rho $ and $\left| {{\psi _n}} \right\rangle $ are the corresponding eigenvectors. Its spin-flipped density matrix $\tilde \rho $ can be expressed as
\begin{align}
\tilde \rho  = {\rm{(}}{\sigma _2} \otimes {\sigma _2}){\rho ^*}{\rm{(}}{\sigma _2} \otimes {\sigma _2}){\rm{ = }}\sum\limits_{n = 1}^4 {{p_n}\left| {{{\tilde \psi }_n}} \right\rangle \left\langle {{{\tilde \psi }_n}} \right|}.
\end{align}
The concurrence is defined by the convex-roof \cite{w18,w19} as follows
\begin{align}
C\left( \rho  \right) = \mathop {\min }\limits_{\left\{ {{q_n},\left| {{\varphi _n}} \right\rangle } \right\}} \sum\limits_n {{q_n}C\left( {\left| {{\varphi _n}} \right\rangle } \right)}.
\end{align}
The minimization is taken over all possible decompositions $\rho $ into pure states. An analytic solution of concurrence can be calculated \cite{w17}
\begin{align}
C\left( \rho  \right) = max\left\{ {0,{\rm{ }}\sqrt {{\lambda _1}}  - \sqrt {{\lambda _2}}  - \sqrt {{\lambda _3}}  - \sqrt {{\lambda _4}} } \right\},
\end{align}
where ${\lambda _n}$ are the eigenvalues, in decreasing order, of the non-Hermitian matrix $\rho \tilde \rho $. The definition of concurrence is based on the convex-roof construction, and it is suitable for use in both pure states and mixed states \cite{w20,w21,w22}.

Besides, a widely used measure of hidden coherence is the first-order coherence \cite{w08}, which is similar with the degree of polarization coherence \cite{w25}. Let us consider a two-qubit state $\rho  = \sum\limits_{n = 1}^4 {{p_n}\left| {{\psi _n}} \right\rangle \left\langle {{\psi _n}} \right|} $, composed of subsystems $A$ and $B$ . This quantum state $\rho $ can be obtained by applying a unitary operation $V$ to the non-entanglement state ${\rho _\Lambda }$, where the unitary operation $V$ contains the corresponding eigenvectors $\left| {{\psi _n}} \right\rangle $ and the state ${\rho _\Lambda }$ is a diagonal matrix with the eigenvalues ${p_n}$. Each subsystem of the state $\rho $ is characterized by the reduced density matrix ${\rho _A} = T{r_B}\left( \rho  \right)$ and ${\rho _B} = T{r_A}\left( \rho  \right)$. The degree of first-order coherence of each subsystem is a better methodology for quantifying this coherence, and it can be given by \cite{w08}
\begin{align}
{D_{A,B}} = \sqrt {2Tr\left( {\rho _{A,B}^2} \right) - 1}.
\end{align}
Therefore, a measure of coherence for both subsystems, when they are considered independently, has the following form \cite{w01}
\begin{align}
D = \sqrt {\frac{{D_A^2 + D_B^2}}{2}}.
\end{align}

\section{A trade-off relation between intrinsic concurrence and coherence}
For a general two-qubit pure state $\left| \psi  \right\rangle $, its concurrence is defined as \cite{w26}
\begin{align}
C\left( {\left| \psi  \right\rangle } \right) = \sqrt {2\left[ {1 - Tr\left( {\rho _B^2} \right)} \right]}  = \sqrt {2\left[ {1 - Tr\left( {\rho _A^2} \right)} \right]},
\end{align}
where ${\rho _A}$ and ${\rho _B}$ are the reduced density matrix of the pure state $\left| \psi  \right\rangle $. Combining the definition of the first-order coherence with the formula (7), it is obvious that the trade-off relation of the pure state $\left| \psi  \right\rangle $ can be expressed as
\begin{align}
{C^2}\left( {\left| \psi  \right\rangle } \right) + {D^2}\left( {\left| \psi  \right\rangle } \right) = 1.
\end{align}
But, for a general two-qubit mixed state, the square sum of these two quantities is not any longer a conserved quantity. It is well-known that if the evolution of the quantum state is unitary, its purity does not change over time. Therefore, the purity can be one of the candidates for conserved quantities. In order to generalize the trade-off relation from the pure state to the mixed state, we try to look for an Hermitian operator whose average value can satisfy the above trade-off relation. Interestingly, we find that the average value of the spin-flipped operator related to a two-qubit mixed state is complementary to its first-order coherence. Here, we introduce some peculiarities about the spin-flipped operator at first.

Let us consider the spin-flipped operator $\tilde F$ corresponding to a Hermitian operator $F$, whose order is $2n$, defined as
\begin{align}
\tilde F = \sigma _2^{ \otimes n}{F^*}\sigma _2^{ \otimes n},
\end{align}
where ${F^*}$ is the complex conjugate of $F$. Obviously, the spin-flipped operator $\tilde F$  satisfies the Hermitian property. If the Hermitian operator $F$ can be written as the form $F = {F_1}{F_2}$, then the corresponding spin-flipped operator $\tilde F$ has a similar form
\begin{align}
\tilde F &= \sigma _2^{ \otimes n}F_1^*F_2^*\sigma _2^{ \otimes n} \nonumber \\&= (\sigma _2^{ \otimes n}F_1^*\sigma _2^{ \otimes n})(\sigma _2^{ \otimes n}F_2^*\sigma _2^{ \otimes n}) \nonumber \\&= {\tilde F_1}{\tilde F_2}.
\end{align}
For the Pauli operators, one obtains some special properties
\begin{align}
{\tilde \sigma _i} = {\sigma _2}\sigma _i^*{\sigma _2} =  - {\sigma _i},
\end{align}
where $i \in \{ 1,{\rm{ }}2,{\rm{ }}3\} $. This indicates that the spin-flipped state ${\tilde \rho _A}$ is related to the single qubit state ${\rho _A}$, is reversed with the state ${\rho _A}$ in the Bloch sphere space. For a general mixed state $\rho  = \sum\limits_{n = 1}^4 {{p_n}\left| {{\psi _n}} \right\rangle \left\langle {{\psi _n}} \right|} $, according to the formula (2), one obtains that the average value of the spin-flipped operator $\tilde \rho $ can be expressed as the following form
\begin{align}
Tr\left( {\rho \tilde \rho } \right) &= \sum\limits_{m,n = 1}^4 {{p_m}{p_n}{C_{mn}}Tr\left( {\left| {{\psi _m}} \right\rangle \left\langle {{{\tilde \psi }_n}} \right|} \right)} \nonumber \\& = \sum\limits_{m,n = 1}^4 {{p_m}{p_n}{C_{mn}}Tr\left( {\left\langle {{{{\tilde \psi }_n}}}
 \mathrel{\left | {\vphantom {{{{\tilde \psi }_n}} {{\psi _m}}}}
 \right. \kern-\nulldelimiterspace}
 {{{\psi _m}}} \right\rangle } \right)} \nonumber \\& = \sum\limits_{m,n = 1}^4 {{p_m}{p_n}{{\left| {{C_{mn}}} \right|}^2}},
\end{align}
where the tilde inner product has the form ${C_{mn}} = \left\langle {{{\psi _m}}}
 \mathrel{\left | {\vphantom {{{\psi _m}} {{{\tilde \psi }_n}}}}
 \right. \kern-\nulldelimiterspace}
 {{{{\tilde \psi }_n}}} \right\rangle $. It is evident that the average $Tr\left( {\rho \tilde \rho } \right)$ is non-negative.
And the tilde inner product ${C_{mn}}$ satisfies some properties
\begin{align}
\sum\limits_{n = 1}^4 {{{\left| {{C_{mn}}} \right|}^2}}  &= \sum\limits_{n = 1}^4 {\left\langle {{{\psi _m}}}
 \mathrel{\left | {\vphantom {{{\psi _m}} {{{\tilde \psi }_n}}}}
 \right. \kern-\nulldelimiterspace}
 {{{{\tilde \psi }_n}}} \right\rangle \left\langle {{{{\tilde \psi }_n}}}
 \mathrel{\left | {\vphantom {{{{\tilde \psi }_n}} {{\psi _m}}}}
 \right. \kern-\nulldelimiterspace}
 {{{\psi _m}}} \right\rangle }  \nonumber \\&= \left\langle {{\psi _m}} \right|\left( {\sum\limits_{n = 1}^4 {\left| {{{\tilde \psi }_n}} \right\rangle \left\langle {{{\tilde \psi }_n}} \right|} } \right)\left| {{\psi _m}} \right\rangle \nonumber \\& {\rm{ = }}\left\langle {{{\psi _m}}}
 \mathrel{\left | {\vphantom {{{\psi _m}} {{\psi _m}}}}
 \right. \kern-\nulldelimiterspace}
 {{{\psi _m}}} \right\rangle  = 1{\rm{ }},
\end{align}
\begin{align}
\sum\limits_{m = 1}^4 {{{\left| {{C_{mn}}} \right|}^2}} & = \sum\limits_{n = 1}^4 {\left\langle {{{{\tilde \psi }_n}}}
 \mathrel{\left | {\vphantom {{{{\tilde \psi }_n}} {{\psi _m}}}}
 \right. \kern-\nulldelimiterspace}
 {{{\psi _m}}} \right\rangle \left\langle {{{\psi _m}}}
 \mathrel{\left | {\vphantom {{{\psi _m}} {{{\tilde \psi }_n}}}}
 \right. \kern-\nulldelimiterspace}
 {{{{\tilde \psi }_n}}} \right\rangle } \nonumber \\& = \left\langle {{{\tilde \psi }_n}} \right|\left( {\sum\limits_{m = 1}^4 {\left| {{\psi _m}} \right\rangle \left\langle {{\psi _m}} \right|} } \right)\left| {{{\tilde \psi }_n}} \right\rangle \nonumber \\& {\rm{ = }}\left\langle {{{{\tilde \psi }_n}}}
 \mathrel{\left | {\vphantom {{{{\tilde \psi }_n}} {{{\tilde \psi }_n}}}}
 \right. \kern-\nulldelimiterspace}
 {{{{\tilde \psi }_n}}} \right\rangle  = 1{\rm{ }}.
\end{align}
And then, we give the definition of the intrinsic concurrence. For a general two-qubit state $\rho $, its intrinsic concurrence is defined as
\begin{align}
{C_I}\left( \rho  \right) = \sqrt {Tr\left( {\rho \tilde \rho } \right)}.
\end{align}
Here, we obtain two properties about the intrinsic concurrence, which indicate the rough relation between the concurrence and the intrinsic concurrence.

Property 1. For a two-qubit pure state $\left| \psi  \right\rangle $, there is an equivalence relation about its concurrence and intrinsic concurrence. The formula can be expressed as
\begin{align}
{C_I}\left( {\left| \psi  \right\rangle } \right) = C\left( {\left| \psi  \right\rangle } \right).
\end{align}

Proof. According to the definition (15), one obtain
\begin{align}
{C_I}\left( {\left| \psi  \right\rangle } \right) & = \sqrt {Tr\left( {\left| \psi  \right\rangle \left\langle {\psi }
 \mathrel{\left | {\vphantom {\psi  {\tilde \psi }}}
 \right. \kern-\nulldelimiterspace}
 {{\tilde \psi }} \right\rangle \left\langle {\tilde \psi } \right|} \right)}  \nonumber \\&= \sqrt {\left\langle {\psi }
 \mathrel{\left | {\vphantom {\psi  {\tilde \psi }}}
 \right. \kern-\nulldelimiterspace}
 {{\tilde \psi }} \right\rangle Tr\left( {\left| \psi  \right\rangle \left\langle {\tilde \psi } \right|} \right)} \nonumber \\&= \sqrt {\left\langle {\psi }
 \mathrel{\left | {\vphantom {\psi  {\tilde \psi }}}
 \right. \kern-\nulldelimiterspace}
 {{\tilde \psi }} \right\rangle Tr\left( {\left\langle {{\tilde \psi }}
 \mathrel{\left | {\vphantom {{\tilde \psi } \psi }}
 \right. \kern-\nulldelimiterspace}
 {\psi } \right\rangle } \right)}\nonumber \\& = \left| {\left\langle {\psi }
 \mathrel{\left | {\vphantom {\psi  {\tilde \psi }}}
 \right. \kern-\nulldelimiterspace}
 {{\tilde \psi }} \right\rangle } \right| = C\left( {\left| \psi  \right\rangle } \right).
\end{align}

Property 2. For a general two-qubit state $\rho $, there is a lower bound of its intrinsic concurrence. And the lower bound is its concurrence. The inequality about its concurrence and intrinsic concurrence can be written as
\begin{align}
{C_I}\left( \rho  \right) \ge C\left( \rho  \right).
\end{align}

Proof. Combining the analytic solution (4) with the definition (15), one obtain
\begin{align}
{C_I}\left( \rho  \right) &= \sqrt {Tr\left( {\rho \tilde \rho } \right)} \nonumber \\& = \sqrt {{\lambda _1} + {\lambda _2} + {\lambda _3} + {\lambda _4}}  \ge \sqrt {{\lambda _1}}  \nonumber \\& \ge max\left\{ {0,\sqrt {{\lambda _1}}  - \sqrt {{\lambda _2}}  - \sqrt {{\lambda _3}}  - \sqrt {{\lambda _4}} } \right\} \nonumber \\&= C\left( \rho  \right).
\end{align}

According to the property 2, one obtain an inference that the necessary and sufficient condition of ${C_I}\left( \rho  \right) = C\left( \rho  \right)$ is $0 \le R\left( {\rho \tilde \rho } \right) \le 1$, where $R\left( {\rho \tilde \rho } \right)$ is the rank of the non-Hermitian matrix $\rho \tilde \rho $. The inference reveals that if a rank-2 mixed state satisfies ${C_I}\left( \rho  \right) = C\left( \rho  \right)$ is $0 \le R\left( {\rho \tilde \rho } \right) \le 1$, the square sum of its first-order coherence and concurrence is a conserved quantity. We will illustrate this for an example in Sec. V.

Finally, we introduce the trade-off relation between the intrinsic concurrence and the first-order coherence. For a general two-qubit state $\rho $, there is a complementary relation
\begin{align}
C_I^2\left( \rho  \right) + {D^2}\left( \rho  \right) = Tr\left( {{\rho ^2}} \right).
\end{align}

Proof. In general, a two-qubit state $\rho $ is denoted as
\begin{align}
\rho  = &\frac{1}{4}[I \otimes I + (\vec A \cdot \vec \sigma ) \otimes I + I \otimes (\vec B \cdot \vec \sigma )\nonumber \\& + \sum\limits_{m,n = 1}^3 {{T_{mn}}{\sigma _m} \otimes {\sigma _n}}],
\end{align}
where $I$ stands for identity operator of single qubit, ${\sigma _n}$ stand for three Pauli operators, $\vec A = \left( {{a_1},{\rm{ }}{a_2},{\rm{ }}{a_3}} \right)$ and $\vec B = \left( {{b_1},{\rm{ }}{b_2},{\rm{ }}{b_3}} \right)$ are vectors in ${R^3}$ and $\vec \sigma  = \left( {{\sigma _1},{\rm{ }}{\sigma _2},{\rm{ }}{\sigma _3}} \right)$. Just to make it easy to calculate, let us rewrite the state $\rho$  as
\begin{align}
\rho {\rm{ = }}{\rho _A} \otimes {\rho _B} + \frac{1}{4}\sum\limits_{m,n = 1}^3 {({T_{mn}} - {a_m}{b_n}){\sigma _m} \otimes {\sigma _n}},
\end{align}
where ${\rho _A} = T{r_B}\left( \rho  \right) = \frac{1}{2}(I + \vec A \cdot \vec \sigma )$ and ${\rho _B} = T{r_A}\left( \rho  \right) = \frac{1}{2}(I + \vec B \cdot \vec \sigma )$. According to the defining (5), we obtain that the coherence of each subsystem has the following form
\begin{align}
{D_K} = \sqrt {2Tr\left( {\rho _K^2} \right) - 1}  = \left| {\vec K} \right|,
\end{align}
where $K \in \left\{ {A,B} \right\}$. Therefore, combining the defining (6) with the formula (23), one obtains that its first-order coherence can be given by the following formula
\begin{align}
D\left( \rho  \right) = \sqrt {\frac{{{{\left| {\vec A} \right|}^2} + {{\left| {\vec B} \right|}^2}}}{2}}.
\end{align}
It is obvious that the spin-flipped operator $\tilde \rho $ corresponding to the state $\rho $  can be expressed as
\begin{align}
\tilde \rho {\rm{ = }}{\tilde \rho _A} \otimes {\tilde \rho _B} + \frac{1}{4}\sum\limits_{m,n = 1}^3 {({T_{mn}} - {a_m}{b_n}){\sigma _m} \otimes {\sigma _n}} ,
\end{align}
where ${\tilde \rho _A} = {\sigma _2}\rho _A^*{\sigma _2} = \frac{1}{2}(I - \vec A \cdot \vec \sigma )$ and ${\tilde \rho _B} = {\sigma _2}\rho _B^*{\sigma _2} = \frac{1}{2}(I - \vec B \cdot \vec \sigma )$. Therefore, for each subsystem, the definition of the first-order coherence can be rewritten as
\begin{align}
{D_K}& = \left| {\vec K} \right| = \sqrt {Tr[{\rho _K}(\vec K \cdot \vec \sigma )]} \nonumber \\& = \sqrt {Tr\left[ {{\rho _K}\left( {{\rho _K} - {{\tilde \rho }_K}} \right)} \right]}.
\end{align}
Similarly, for the whole system, the definition of the first-order coherence can be rewritten as
\begin{align}
D\left( \rho  \right)& = \sqrt {\frac{{{{\left| {\vec A} \right|}^2} + {{\left| {\vec B} \right|}^2}}}{2}}  \nonumber \\&= \sqrt {\frac{{Tr\left[ {\rho (\vec A \cdot \vec \sigma  \otimes I)} \right] + Tr\left[ {\rho (I \otimes \vec B \cdot \vec \sigma )} \right]}}{2}}  \nonumber \\&= \sqrt {Tr\left[ {\rho ({\rho _A} \otimes {\rho _B} - {{\tilde \rho }_A} \otimes {{\tilde \rho }_B})} \right]}.
\end{align}
Therefore, the square sum of its first-order coherence and intrinsic concurrence is a conserved quantity. The derivative process can be described as follows
\begin{align}
C_I^2\left( \rho  \right) + {D^2}\left( \rho  \right) &= Tr\left( {\rho \tilde \rho } \right) + Tr\left[ {\rho ({\rho _A} \otimes {\rho _B} - {{\tilde \rho }_A} \otimes {{\tilde \rho }_B})} \right] \nonumber \\&= Tr\left[ {\rho ({\rho _A} \otimes {\rho _B} - {{\tilde \rho }_A} \otimes {{\tilde \rho }_B} + \tilde \rho )} \right]\nonumber \\& = Tr\left( {{\rho ^2}} \right).
\end{align}

Note that, the trade-off relation reveals that for a general two-qubit state $\rho $, its first-order coherence $D\left( \rho  \right)$ and intrinsic concurrence ${C_I}\left( \rho  \right)$ are a pair of complementary quantities. And it shows that if the evolution of the whole system is unitary, there is a mutual transformation relationship between its first-order coherence and intrinsic concurrence. In theory, for a general two-qubit state $\rho  = V{\rho _\Lambda }{V^\dag }$, we can always apply a unitary operation $U = M{V^\dag }$ to it to get a bell diagonal state ${\rho _{BDS}} = M{\rho _\Lambda }{M^\dag }$, where the matrix
\begin{align}
M = \frac{1}{{\sqrt 2 }}\left( {\begin{array}{*{20}{c}}
1&1&0&0\\
0&0&1&1\\
0&0&1&{ - 1}\\
1&{ - 1}&0&0
\end{array}} \right).
\end{align}
And the state ${\rho _{BDS}}$ with minimal first-order coherence $D\left( {{\rho _{BDS}}} \right) = 0$  maximizes the intrinsic concurrence with the value
\begin{align}
{C_I}\left( {{\rho _{BDS}}} \right) = \sqrt {Tr\left( {{\rho ^2}} \right)}.
\end{align}
In the same way, we can always apply a unitary operation $U = {V^\dag }$ to the state $\rho  = V{\rho _\Lambda }{V^\dag }$ to get a non-entanglement state ${\rho _\Lambda }$. And the state ${\rho _\Lambda }$ with maximal first-order coherence $D\left( {{\rho _\Lambda }} \right) = \sqrt {{{\left( {{p_1} - {p_4}} \right)}^2} + {{\left( {{p_2} - {p_3}} \right)}^2}} $  minimizes the intrinsic concurrence with the value
\begin{align}
{C_I}\left( {{\rho _\Lambda }} \right) = \sqrt {2\left( {{p_1}{\rm{ }}{p_4} + {p_2}{\rm{ }}{p_3}} \right)}.
\end{align}

\begin{table*}\vskip 0.2cm\
\caption{The unitary evolution operators ${U_n}$ and the evolution states ${\rho _n}$ by these channels respectively.}
\begin{tabular}{|c|c|c|}
\hline
       Channels & AD & BF   \\
\hline
  Unitary evolution operators ${U_n}$
&${U_0} = \left( {\begin{array}{*{20}{c}}
1&0&0&0\\
0&{\sqrt {1 - p} }&{\sqrt p }&0\\
0&{ - \sqrt p }&{\sqrt {1 - p} }&0\\
0&0&0&1
\end{array}} \right)$
& ${U_1} = \left( {\begin{array}{*{20}{c}}
{\sqrt {1 - p} }&0&0&{ - \sqrt p }\\
0&{\sqrt {1 - p} }&{\sqrt p }&0\\
0&{ - \sqrt p }&{\sqrt {1 - p} }&0\\
{\sqrt p }&0&0&{\sqrt {1 - p} }
\end{array}} \right)$  \\

\hline
  Evolution states ${\rho _n}$
   & ${\rho _0} = {U_0}{\rho _{IT}}U_0^\dag $
   & ${\rho _1} = {U_1}{\rho _{IT}}U_1^\dag $  \\

 \hline
       Channels & BPF & PF   \\
\hline
  Unitary evolution operators ${U_n}$
&${U_2} = \left( {\begin{array}{*{20}{c}}
{\sqrt {1 - p} }&0&0&{i\sqrt p }\\
0&{\sqrt {1 - p} }&{ - i\sqrt p }&0\\
0&{ - i\sqrt p }&{\sqrt {1 - p} }&0\\
{i\sqrt p }&0&0&{\sqrt {1 - p} }
\end{array}} \right)$
& ${U_3} = \left( {\begin{array}{*{20}{c}}
{\sqrt {1 - p} }&{ - \sqrt p }&0&0\\
{\sqrt p }&{\sqrt {1 - p} }&0&0\\
0&0&{\sqrt {1 - p} }&{\sqrt p }\\
0&0&{ - \sqrt p }&{\sqrt {1 - p} }
\end{array}} \right)$ \\

\hline
  Evolution states ${\rho _n}$
   & ${\rho _2} = {U_2}{\rho _{IT}}U_2^\dag $
   & ${\rho _3} = {U_3}{\rho _{IT}}U_3^\dag $ \\
\hline
\end{tabular}
\end{table*}

\begin{table*}
\caption{The ranks ${R_n}$ and ratios ${S_n}$ of the evolution states ${\rho _n}$ by these channels respectively }
\begin{tabular}{|c|c|c|c|c|}
\hline
       Channels & AD & BF  & BPF & PF \\
\hline
  Ranks ${R_n}$
&${R_0} = 1$
& ${R_1} = 2$
& ${R_2} = 2$
& ${R_3} = 2$ \\

\hline
  Ratios ${S_n}$
   & ${S_0} = 1$
   & ${S_1} = \sqrt {\frac{{1{\rm{ + }}{{\left| {\vec A} \right|}^2} - 2a_1^2}}{{2\left( {{{\left| {\vec A} \right|}^2} - a_1^2} \right)}}}$
   & ${S_2} = \sqrt {\frac{{1{\rm{ + }}{{\left| {\vec A} \right|}^2} - 2a_2^2}}{{2\left( {{{\left| {\vec A} \right|}^2} - a_2^2} \right)}}}$
   & ${S_3} = \sqrt {\frac{{1{\rm{ + }}{{\left| {\vec A} \right|}^2} - 2a_3^2}}{{2\left( {{{\left| {\vec A} \right|}^2} - a_3^2} \right)}}}$ \\
\hline
\end{tabular}
\end{table*}

\begin{figure}
\centering
\includegraphics[width=7cm,height=5cm]{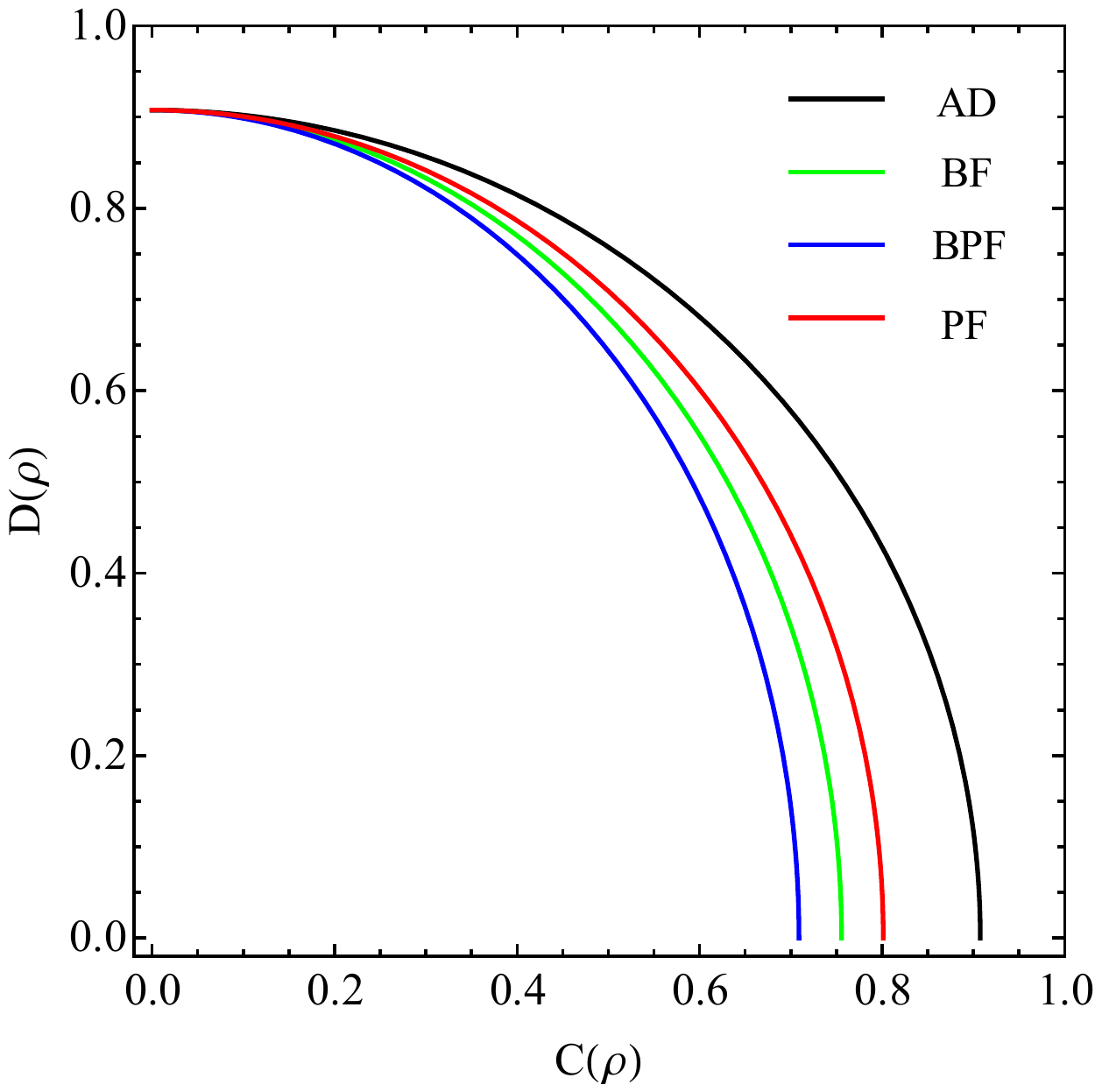}
\caption{ The first-order coherence $D\left( \rho  \right)$ versus the concurrence $C\left( \rho  \right)$ for the composite system $\rho $, which is formed by the single qubit state ${\rho _A} = \frac{1}{2}(I + \vec A \cdot \vec \sigma )$ and four coupling channels respectively, where $\vec A = \left( {0.50,{\rm{ }}0.61,{\rm{ }}0.16} \right)$.}
\label{g0}
\end{figure}

\section{The relation between the intrinsic concurrence and the concurrence for a two-qubit state}
For a two-qubit state $\rho $, it can be decomposed according to the eigenvectors of the non-Hermitian matrix $\rho \tilde \rho $. And we give a relationship about the intrinsic concurrence with the concurrence of a special pure state ensemble. For that, let's prove the following theorems.

Theorem 1. If a two-qubit state $\rho $ has a pure state decomposition $\rho  = \sum\limits_{n = 1}^4 {{q_n}\left| {{\varphi _n}} \right\rangle \left\langle {{\varphi _n}} \right|} $, and these pure states $\left| {{\varphi _n}} \right\rangle $ satisfy the tilde orthogonal relation $\left\langle {{{\varphi _m}}}
 \mathrel{\left | {\vphantom {{{\varphi _m}} {{{\tilde \varphi }_n}}}}
 \right. \kern-\nulldelimiterspace}
 {{{{\tilde \varphi }_n}}} \right\rangle  = {\delta _{mn}}\left\langle {{{\varphi _n}}}
 \mathrel{\left | {\vphantom {{{\varphi _n}} {{{\tilde \varphi }_n}}}}
 \right. \kern-\nulldelimiterspace}
 {{{{\tilde \varphi }_n}}} \right\rangle $, then the eigenvectors of the non-Hermitian matrix $\rho \tilde \rho $ will be $\left| {{\varphi _n}} \right\rangle $ and the corresponding eigenvalues will be expressed as ${\lambda _{\rm{n}}} = q_n^2{\rm{ }}{C^2}\left( {\left| {{\varphi _n}} \right\rangle } \right)$.

 Proof. Assuming that the two-qubit state $\rho $ has a pure state decomposition
\begin{align}
\rho  = \sum\limits_{n = 1}^4 {{q_n}\left| {{\varphi _n}} \right\rangle \left\langle {{\varphi _n}} \right|},
\end{align}
where these pure states $\left| {{\varphi _n}} \right\rangle $ satisfy the tilde orthogonal relation
\begin{align}
\left\langle {{{\varphi _m}}}
 \mathrel{\left | {\vphantom {{{\varphi _m}} {{{\tilde \varphi }_n}}}}
 \right. \kern-\nulldelimiterspace}
 {{{{\tilde \varphi }_n}}} \right\rangle  = {\delta _{mn}}\left\langle {{{\varphi _n}}}
 \mathrel{\left | {\vphantom {{{\varphi _n}} {{{\tilde \varphi }_n}}}}
 \right. \kern-\nulldelimiterspace}
 {{{{\tilde \varphi }_n}}} \right\rangle.
\end{align}
And then, combining the formula (32) with (33), we obtain that the non-Hermitian matrix $\rho \tilde \rho $  can be expressed as
\begin{align}
\rho \tilde \rho & = \left( {\sum\limits_{m = 1}^4 {{q_m}\left| {{\varphi _m}} \right\rangle \left\langle {{\varphi _m}} \right|} } \right)\left( {\sum\limits_{n = 1}^4 {{q_n}\left| {{{\tilde \varphi }_n}} \right\rangle \left\langle {{{\tilde \varphi }_n}} \right|} } \right) \nonumber \\&= \sum\limits_{m = 1}^4 {q_m^2\left\langle {{{\varphi _m}}}
 \mathrel{\left | {\vphantom {{{\varphi _m}} {{{\tilde \varphi }_m}}}}
 \right. \kern-\nulldelimiterspace}
 {{{{\tilde \varphi }_m}}} \right\rangle \left| {{\varphi _m}} \right\rangle \left\langle {{{\tilde \varphi }_m}} \right|}.
\end{align}
So, one obtain that the eigenvalue-eigenvector equation of the non-Hermitian matrix $\rho \tilde \rho $  has the form
\begin{align}
\rho \tilde \rho \left| {{\varphi _n}} \right\rangle  &= \sum\limits_{m = 1}^4 {q_m^2\left\langle {{{\varphi _m}}}
 \mathrel{\left | {\vphantom {{{\varphi _m}} {{{\tilde \varphi }_m}}}}
 \right. \kern-\nulldelimiterspace}
 {{{{\tilde \varphi }_m}}} \right\rangle \left| {{\varphi _m}} \right\rangle \left\langle {{{{\tilde \varphi }_m}}}
 \mathrel{\left | {\vphantom {{{{\tilde \varphi }_m}} {{\varphi _n}}}}
 \right. \kern-\nulldelimiterspace}
 {{{\varphi _n}}} \right\rangle } \nonumber \\& = q_n^2{\left| {\left\langle {{{\varphi _n}}}
 \mathrel{\left | {\vphantom {{{\varphi _n}} {{{\tilde \varphi }_n}}}}
 \right. \kern-\nulldelimiterspace}
 {{{{\tilde \varphi }_n}}} \right\rangle } \right|^2}\left| {{\varphi _n}} \right\rangle  \nonumber \\&= q_n^2{C^2}\left( {\left| {{\varphi _n}} \right\rangle } \right)\left| {{\varphi _n}} \right\rangle.
\end{align}

Theorem 2. For a general two-qubit state $\rho $, if the eigenvalue-eigenvector equation of the non-Hermitian matrix $\rho \tilde \rho $ has the form $\rho \tilde \rho \left| {{\varphi _n}} \right\rangle  = {\lambda _n}\left| {{\varphi _n}} \right\rangle $, these eigenvectors $\left| {{\varphi _n}} \right\rangle $ will satisfy the tilde orthogonal relation $\left\langle {{{\varphi _m}}}
 \mathrel{\left | {\vphantom {{{\varphi _m}} {{{\tilde \varphi }_n}}}}
 \right. \kern-\nulldelimiterspace}
 {{{{\tilde \varphi }_n}}} \right\rangle  = {\delta _{mn}}\left\langle {{{\varphi _n}}}
 \mathrel{\left | {\vphantom {{{\varphi _n}} {{{\tilde \varphi }_n}}}}
 \right. \kern-\nulldelimiterspace}
 {{{{\tilde \varphi }_n}}} \right\rangle $ and the state $\rho $ will have a pure state decomposition $\rho  = \sum\limits_{n = 1}^4 {{q_n}\left| {{\varphi _n}} \right\rangle \left\langle {{\varphi _n}} \right|} $, where ${q_n}$ satisfy the relation ${\lambda _{\rm{n}}} = q_n^2{\rm{ }}{C^2}\left( {\left| {{\varphi _n}} \right\rangle } \right)$.

 Proof. The eigenvalue-eigenvector equation of the non-Hermitian matrix $\rho \tilde \rho $ about the state $\rho $ can be expressed as
\begin{align}
\rho \tilde \rho \left| {{\varphi _n}} \right\rangle  = {\lambda _n}\left| {{\varphi _n}} \right\rangle.
\end{align}
And then let's apply spin-flip to both sides of the equation (36), one can obtain
\begin{align}
\tilde \rho \rho \left| {{{\tilde \varphi }_n}} \right\rangle  = {\lambda _n}\left| {{{\tilde \varphi }_n}} \right\rangle.
\end{align}
It is obvious that the non-Hermitian matrix $\tilde \rho \rho $ is the Hermitian conjugate of the non-Hermitian matrix $\rho \tilde \rho $. Therefore, we obtain that the eigenvalue spectral decomposition of the non-Hermitian matrix $\rho \tilde \rho $ can be expressed as \cite{w27}
\begin{align}
\rho \tilde \rho  = \sum\limits_{n = 1}^4 {\frac{{{\lambda _n}}}{{\left\langle {{{{\tilde \varphi }_n}}}
 \mathrel{\left | {\vphantom {{{{\tilde \varphi }_n}} {{\varphi _n}}}}
 \right. \kern-\nulldelimiterspace}
 {{{\varphi _n}}} \right\rangle }}\left| {{\varphi _n}} \right\rangle \left\langle {{{\tilde \varphi }_n}} \right|},
\end{align}
and these eigenvectors $\left| {{\varphi _n}} \right\rangle $ satisfy the tilde orthogonal relation $\left\langle {{{\varphi _m}}}
 \mathrel{\left | {\vphantom {{{\varphi _m}} {{{\tilde \varphi }_n}}}}
 \right. \kern-\nulldelimiterspace}
 {{{{\tilde \varphi }_n}}} \right\rangle  = {\delta _{mn}}\left\langle {{{\varphi _n}}}
 \mathrel{\left | {\vphantom {{{\varphi _n}} {{{\tilde \varphi }_n}}}}
 \right. \kern-\nulldelimiterspace}
 {{{{\tilde \varphi }_n}}} \right\rangle $.
And then there is a special decomposition of the state $\rho $
\begin{align}
\rho  = \sum\limits_{n = 1}^4 {{q_n}\left| {{\varphi _n}} \right\rangle \left\langle {{\varphi _n}} \right|},
\end{align}
where ${q_n}$ satisfy the relation
\begin{align}
{\lambda _{\rm{n}}} = q_n^2{\rm{ }}{C^2}\left( {\left| {{\varphi _n}} \right\rangle } \right).
\end{align}

In fact, the theory 2 provides a special pure state decomposition method for a two-qubit state $\rho $, which will provide a convenience for us to solve the pure state decomposition. According to the relation (40), one can get a relation between the intrinsic concurrence and the concurrence of the special pure state ensemble, i.e.
\begin{align}
C_I^2\left( \rho  \right) = \sum\limits_{n = 1}^4 {q_n^2{\rm{ }}{C^2}\left( {\left| {{\varphi _n}} \right\rangle } \right)}.
\end{align}
The relationship (41) reveals that the intrinsic concurrence ${C_I}\left( \rho  \right)$ is inseparable from the concurrence $C\left( {\left| {{\varphi _n}} \right\rangle } \right)$ of the special pure state ensemble about the state $\rho $.

\section{The complementary relation of quantum state in open system}
In fact, a quantum system is inevitably coupled with the surrounding environment, and quantum resources are constantly exchanged between the quantum system and the environment. In general, in order to explore the evolution of a single qubit state ${\rho _A} = \frac{1}{2}(I + \vec A \cdot \vec \sigma )$ in open system, one can be simplified into a closed composite system by Markovian approximation. We assume that the initial state of the environment is $\left| 0 \right\rangle \left\langle 0 \right|$. Then the state of the composite system at the initial time (IT) can be described as ${\rho _{IT}} = {\rho _A} \otimes \left| 0 \right\rangle \left\langle 0 \right|$. And its evolution state $\rho $ can be described by unitary evolution operator $U$ , i.e. $\rho  = U{\rho _{IT}}{U^\dag }$.

We will discuss only a few common channels, namely amplitude damped (AD) channel, bit flip (BF) channel, bit-phase flip (BPF) channel, and phase flip (PF) channel. The unitary evolution operators ${U_n}$ and the evolution states ${\rho _n}$, which formed by the coupling of the single qubit state ${\rho _A}$ and these channels respectively, are given in the TABLE I. For these channels, we obtain that the ratios ${S_n} = \frac{{{C_I}\left( {{\rho _n}} \right)}}{{C\left( {{\rho _n}} \right)}}$ are independent of the parameter ${p}$ of the channels. The ranks ${R_n}$ of the non-Hermitian matrix ${\rho _n}{\tilde \rho _n}$ and the ratios ${S_n}$ are given in the TABLE II.

Therefore, the complementary relation of the composite system ${\rho _n}$, which is formed by the single qubit state ${\rho _A}$ and its coupling channel, can be rewritten as
\begin{align}
{D^2}\left( {{\rho _n}} \right) + S_n^2{\rm{ }}{C^2}\left( {{\rho _n}} \right) = \frac{{1 + {{\left| {\vec A} \right|}^2}}}{2},
\end{align}
where $n \in \{ 0,1,2,3\} $.
The above equation (42) expresses the following two meanings: one is that when a single qubit state ${\rho _A}$ is coupled with the AD channel, the relation between the concurrence and first-order coherence can be revealed by using a circular curve in Fig.1 and the concurrence of the composite system will be increased completely from the decrease of its first-order coherence; the other is that when the state ${\rho _A}$ is coupled with the BF, BPF and PF channels respectively, the relationship between the concurrence and first-order coherence can be represented by an elliptic curve in Fig.1 and the decrease of the first-order coherence can be converted to the concurrence with a conversion efficiency $\frac{1}{{{S_n}}}$.

\section{Conclusion}
In this paper, we have solved three tasks about the trade-off relation between intrinsic concurrence and first-order coherence. First of all, we put forward the definition of intrinsic concurrence for a general two-qubit state and establish a universal relation that its intrinsic concurrence can be complementary to its first-order coherence. Then, we provide a special decomposition method for a two-qubit state and derive the relation between its concurrence and intrinsic concurrence. Finally, as an application, we give out the unified complementary relation of single-qubit state under four different noise channels. Our results provide a deep physical meaning about the relation between quantum coherence and entanglement which can be widely applied in various contexts. Quantum entanglement of the compound system will be increased completely from the decrease of quantum coherence through an amplitude damped channel. But If quantum systems pass through others channels, the decrease of quantum coherence can be converted to quantum entanglement with a conversion efficiency. We hope that these results will find interesting applications in controlling interconversion of quantum coherence and entanglement for a two-qubit system and sending a single-qubit system under noisy channels that tend to enhance quantum coherence or entanglement.	

\section*{Acknowledgement} 
	This work was supported by the National Science Foundation of China under Grant Nos. 11575001 and 61601002, Anhui Provincial Natural Science Foundation (Grant No. 1508085QF139) and Natural Science Foundation of Education Department of Anhui Province (Grant No. KJ2016SD49), and also the fund from CAS Key Laboratory of Quantum Information (Grant No. KQI201701).

\bibliographystyle{plain}

\end{document}